\documentclass[useAMS,usenatbib,usegraphicx]{mn2e}

\usepackage[dvips,dvipdfm]{graphicx,color}
\usepackage[dvips,dvipdfm]{graphicx}
\usepackage{aas_macros}
\definecolor{mygray}{gray}{0.7}

\title[Size growth of cluster ETGs]{Mass and size growth of early-type galaxies by dry mergers  in cluster environments}
\author[T. Oogi, A. Habe and T. Ishiyama]{Taira Oogi,$^{1,2,3,4}$\thanks{E-mail: oogi@koshigaya.bunkyo.ac.jp (TO)}, Asao Habe$^{1}$\thanks{E-mail: habe@astro1.sci.hokudai.ac.jp (AH)} and Tomoaki Ishiyama$^{5,6}$\thanks{E-mail: ishiyama@chiba-u.jp (TI)} \\
$^{1}$Department of Cosmosciences, Graduate School of Science, Hokkaido University, Kita-10 Nishi-8, Kita-ku, Sapporo 060-0810, Japan \\
$^{2}$Center for Computational Astrophysics, National Astronomical Observatory of Japan, 2-21-1, Osawa, Mitaka City, Tokyo 181-8588, Japan \\
$^{3}$Faculty of Education, Nagasaki University, 1-14, Bunkyo-machi, Nagasaki City, Nagasaki 852-8521, Japan \\
$^{4}$Faculty of Education, Bunkyo University, 3337 Minami-Ogishima, Koshigaya-shi, Saitama 343-8511, Japan \\
$^{5}$Center for Computational Science, University of Tsukuba, Tsukuba, 1-1-1 Tennodai, Tsukuba, Ibaraki 305-8577, Japan \\
$^{6}$Institute of Management and Information Technologies, Chiba University, 1-33, Yayoi-cho, Inage-ku, Chiba 263-8522, Japan}

\begin{document}

\date{}

\pagerange{\pageref{firstpage}--\pageref{lastpage}} \pubyear{2015}

\maketitle

\label{firstpage}

\begin{abstract}
We perform dry merger simulations to investigate the role of dry mergers in the size growth of early-type galaxies in high density environments.
We replace the virialized dark matter haloes obtained by a large cosmological $N$-body simulation with $N$-body galaxy models consisting of two components, a stellar bulge and a dark matter halo, which have higher mass resolution than the cosmological simulation.
We then re-simulate nine cluster forming regions, whose masses range from 1$\times 10^{14}\mathrm{M}_{\odot}$ to 5$\times 10^{14}\mathrm{M}_{\odot}$.
Masses and sizes of stellar bulges are also assumed to satisfy the stellar mass--size relation of high-z compact massive early-type galaxies.
We find that dry major mergers considerably contribute to the mass and size growth of central massive galaxies.
One or two dry major mergers double the average stellar mass and quadruple the average size between $z=2$ and $z=0$.
These growths favorably agree with observations.
Moreover, the density distributions of our simulated central massive galaxies grow from the inside-out, which is consistent with recent observations.
The mass--size evolution is approximated as $ R\propto M_{*}^{\alpha}$, with $\alpha \sim 2.24$.
Most of our simulated galaxies are efficiently grown by dry mergers, and their stellar mass--size relations match the ones observed in the local Universe.
Our results show that the central galaxies in the cluster haloes are potential descendants of high-z ($z\sim$ 2-3) compact massive early-type galaxies.
This conclusion is consistent with previous numerical studies which investigate the formation and evolution of compact massive early-type galaxies.
\end{abstract}


\section{Introduction}

Observations strongly suggest that massive ($\ga 10^{11} \mathrm{M}_{\odot}$) early-type galaxies (ETGs) at high-z ($z\sim 2 - 3$) are more compact than galaxies with comparable mass in the local Universe.
This finding challenges the $\Lambda$ cold dark matter ($\Lambda$CDM) model of the structure formation of the Universe.
High-z ETGs are 3-5 times smaller than local ETGs with comparable mass (e.g. \citealt{2005ApJ...626..680D}, \citealt{2006ApJ...650...18T}, \citealt{2007MNRAS.382..109T}, \citealt{2008ApJ...687L..61B}, \citealt{2008A&A...482...21C}, \citealt{2008ApJ...677L...5V}).
These high-z ETGs also lie below the stellar mass-size relation in the local Universe, a relation well investigated through the Sloan Digital Sky Survey (SDSS) database (e.g. \citealt{2003MNRAS.343..978S}).
The stellar mass densities of these high-z ETGs are inferred to be 1-2 orders of magnitude higher than those of local ETGs.
Several recent observations have also revealed higher velocity dispersions of high-z ETGs than those of local ETGs with comparable mass (\citealt{2009ApJ...696L..43C}, \citealt{2009ApJ...704L..34C}, \citealt*{2009Natur.460..717V}, \citealt{2011ApJ...736L...9V}).
Compact massive ETGs at high-z exhibit quiescent star formation activity and relatively old stellar populations (e.g. \citealt{2008A&A...482...21C}, \citealt{2008ApJ...677L...5V}), similarly to ETGs in the local Universe (e.g. \citealt{2005ApJ...621..673T}).
Moreover, they are extremely rare in the local Universe (e.g. \citealt{2010ApJ...720..723T}).
Because local ETGs also contain old stellar populations, it appears that compact massive ETGs at high-z must increase their size and decrease their velocity dispersion from $z\simeq 2$ to $z=0$ without recent star formation.
This problem is called the `size evolution problem of ETGs'.

An ETG can evolve its size without star formation if it undergoes a dry merger event (e.g. \citealt{2006ApJ...650..791C}; \citealt*{2007ApJ...658...65C}; \citealt{2013MNRAS.428..641O}, hereafter \mbox{paper I}).
Dry mergers can increase a galaxy's size more than wet (i.e. gas-rich) mergers because of the absence of energy dissipation via the gas component. 
In addition, dry minor mergers are a more efficient process for the size evolution of ETGs than dry major mergers (\citealt{2009ApJ...697.1290B}; \citealt*{2009ApJ...699L.178N}; paper I).
In paper I, we have demonstrated by $N$-body simulations that sequential minor mergers of compact satellite galaxies most effectively promote size growth and decrease the velocity dispersion of compact massive ETGs.
The role of dry mergers for the formation of ETGs is also discussed in 
terms of the inner surface brightness profiles of local ETGs (\citealt{2009ApJS..182..216K}; \citealt{2012ApJS..198....2K}).

It is interesting to study dry merger processes in the context of the $\Lambda$CDM model.
In paper I, we have analysed the Millennium Simulation Data Base (\citealt{2005Natur.435..629S}; \citealt{2007MNRAS.375....2D}), and shown that the mass evolution of ETGs is mainly governed by dry major and minor mergers.
The study of mergers in high density environments is especially interesting because ETGs are more common in high density environments (galaxy groups and clusters) than in low density environments (e.g. \citealt{1980ApJ...236..351D}) in the local Universe.
Using a semi-analytic model, \citet{2013MNRAS.428..109S} have reported that the size evolution of ETGs largely depends on the galactic environment.
They have also predicted that size evolution in galaxies inhabiting massive haloes is strengthened by the large number of mergers (see also paper I).
We may naturally expect that many high-z compact massive ETGs evolved into the ETGs observed in present high density environments.
In such dense environments, the size evolution of ETGs may be enhanced by dry mergers.
According to recent observations, major mergers have strongly influenced the mass growth of massive ETGs (\citealt{2012A&A...548A...7L}; \citealt{2014MNRAS.444..906F}) and the brightest cluster galaxies (\citealt{2013MNRAS.433..825L}; \citealt{2013MNRAS.434.2856B}).

There are some numerical studies to investigate the effects of the merging process in high density environments (\citealt{1998ApJ...502..141D}; \citealt*{2006ApJ...648..936R}; \citealt{2007MNRAS.377....2M}; \citealt{2010MNRAS.406..936P}; \citealt*{2011ApJ...732...48R}; \citealt{2012MNRAS.424..747L}), however, most of these have not focused on the size growth of compact massive ETGs.
\citet{2009ApJ...699L.178N} and \citet{2012ApJ...744...63O} performed cosmological hydrodynamic simulations of massive early-type galaxy formation.
Both the studies showed that, in the field environments, ETGs increase sizes from $z\sim3$ to $z=0$ through dry minor mergers.
However, neither study explored the size growth in high density environments.

Indeed, few numerical studies have simulated the dry merging process in high density environments (\citealt{2009ApJ...696.1094R}, \citealt{2013MNRAS.435..901L}).
\citet{2009ApJ...696.1094R} assumed a constant stellar to dark halo mass ratio, $M_{*} / M_{\mathrm{halo}} = 1/9$, for all of their model galaxies at $z=3$.
This value is significantly larger than that reported in recent abundance matching studies of the observed high-z galaxies (\citealt{2010ApJ...710..903M}; \citealt{2010MNRAS.404.1111G}; \citealt*{2010ApJ...717..379B}).
The authors also assumed the local stellar mass--size relation given by \citet{2003MNRAS.343..978S}.
However, this assumption cannot be extended to high-z massive galaxies, which are highly compact as discussed above.
\citet{2012MNRAS.424..747L} also pointed out the above shortcomings of the previous studies.
\citet{2013MNRAS.435..901L} analysed nine high-resolution cosmological dark matter simulations of galaxy clusters from the Phoenix project (\citealt{2012MNRAS.425.2169G}).
They investigated the size and mass growths of cluster galaxies by assigning star particles to the original dark matter particles on the basis of the stellar-to-halo relation obtained from the abundance matching method.
In their approach, they used a weighting scheme, and stellar systems were not separated from dark matter haloes.
However, stellar systems may evolve as different collisionless systems from dark matter haloes because stars are formed in dense gas within the deep gravitational potential of dark matter haloes.
In addition, \citet{2013MNRAS.435..901L} focused on the growth of the brightest cluster galaxies in rich clusters, whose masses exceed $5\times10^{14} h^{-1} \mathrm{M}_{\odot}$.

In this paper, we investigate the mass and size growth of high-z compact massive ETGs in dense environments by dry merging process.
The ETGs are evolved up to the present Universe.
Dry mergers of the high-z galaxies are synthesized in cosmological $N$-body simulations by the re-simulation technique described in the next section.
In our simulations, massive galaxies are grown in less massive clusters ($10^{14} \mathrm{M}_{\odot} \la M \la 5\times10^{14} \mathrm{M}_{\odot}$ at $z=0$).
This mass range was not covered in previous studies by by \citet{2012ApJ...744...63O} and \citet{2013MNRAS.435..901L}.
In the dry merger scenario, the efficiency of size growth depends on the stellar-to-halo mass ratio of the galaxy.
Here, we assume the stellar-to-halo mass relation derived from abundance matching studies of high-z galaxies (\citealt{2010ApJ...710..903M}, see also \citealt{2010MNRAS.404.1111G} and \citealt{2010ApJ...717..379B})
\footnote{More recently, \citet*{2013MNRAS.428.3121M} and \citet*{2013ApJ...770...57B} also derive the stellar-to-mass relations, respectively.
At $z\sim3$, the stellar masses in haloes with $10^{13} M_{\odot}$ in all studies show agreement with each other, that is $M_{*}\sim10^{11} M_{\odot}$.
On the other hand, at low halo mass end ($M_{\mathrm{halo}} \sim 10^{11} M_{\odot}$), the results of \citet{2013MNRAS.428.3121M} and \citet{2013ApJ...770...57B} predict larger stellar mass than that of \citet{2010ApJ...710..903M}, which we adopted in this paper.
However, the relation of \citet{2010ApJ...710..903M} is within the plausibility range of \citet{2013MNRAS.428.3121M}.
Thus, our assumption for initial stellar masses is consistent with those papers.}.
We investigate the relative importance of dry major and minor mergers for the mass and size growth of compact massive ETGs in clusters.
Further, we compare our numerical results of the growth with the observed stellar mass--size relations in the local Universe.

This remainder of this paper is organized as follows.
In \S 2, we outline the simulations, presenting our cosmological $N$-body simulation, initial models and selected simulation parameters.
The merger remnants of our simulations are analysed and the results presented in \S 3.
In \S 4, we summarize the paper and discuss the evolutionary paths of our simulated galaxies, and compare our results with those of previous theoretical and observational studies.

\section{Simulations}

We investigate the mass and size evolutions of cluster galaxies similar to \citet{1998ApJ...502..141D}, \citet{2006ApJ...648..936R} and \citet{2009ApJ...696.1094R}.
We simply assume that spheroidal galaxies form at the centres of dark matter haloes at $z\sim3$, the epoch of observed compact massive ETGs.
The ETGs at $z\sim3$ are simulated in equilibrium $N$-body galaxy models with higher numerical resolution than our cosmological dark matter simulation.
The stellar components of the model galaxies are assumed from the stellar-to-halo mass relations obtained via abundance matching by \citet{2010ApJ...710..903M}.
Cluster regions are simulated by the following procedure:

\begin{enumerate}
\item We perform a large scale cosmological dark matter simulation (see \S\ref{cosmo_sim}) until $z=0$.
\item We randomly select nine cluster-sized haloes (hereafter, called the {\it cluster haloes}) from $\sim50$ cluster haloes with virial masses $M_{\mathrm{vir}} \simeq 10^{14} - 5\times 10^{14} \mathrm{M}_{\odot}$ yielded by the cosmological simulation at $z=0$.
\item The particles within twice the virial radius of each selected cluster halo are traced back to $z=2.85$.
\item We identify the virialized dark matter haloes with masses exceeding $10^{11} \mathrm{M}_{\odot}$ at $z=2.85$, and replace these haloes with our galaxy models, which have higher mass resolution (see \S\ref{gal_model}).
\item The cluster region is re-simulated until $z=0$ (see \S\ref{re_sim}).
\end{enumerate}
This procedure is implemented for the nine randomly selected cluster regions.
Steps (i)--(v) are described in more detail below.

\subsection{Cosmological dark matter simulation} \label{cosmo_sim}

The cluster haloes were selected in a high-resolution cosmological simulation using the GreeM code (\citealt*{2009PASJ...61.1319I}; \citealt*{Ishiyama2012}).
GreeM is a massively parallel TreePM code suitable  for very large cosmological $N$-body simulations.
The cosmological parameters were based on the $\Lambda$CDM cosmological model ($\Omega_0=0.3, \lambda_0=0.7, h=0.7, \sigma_8=0.8, n=1.0$), which is consistent with the observational results obtained by the Planck satellite (\citealt{2014A&A...571A..16P}).
The similar parameter values were adopted in our previous simulation (\citealt{2013ApJ...767..146I}; \citealt{2015PASJ...67...61I}).
We constructed a cube with a comoving size of 171.4 Mpc and periodic boundary conditions.
The cube contained $1600^3$ particles, corresponding to a mass resolution of $5.0\times10^7\mathrm{M}_{\odot}$.
The initial redshift was 83.
The gravitational Plummer softening length was 2.6kpc at $z=0$.
The softening was constant in the comoving coordinates from $z=83$ (initial condition) to $z=4$.
From $z=4$ to $z=0$, it was constant in the physical coordinates.
The initial particle distributions were generated by the MPGRAFIC package (\citealt{2008ApJS..178..179P}), which is a parallelized variation of the GRAFIC package (\citealt{2001ApJS..137....1B}).
More than 50 cluster haloes with virial masses exceeding $10^{14}\mathrm{M}_{\odot}$ were identified in the cosmological simulation at $z=0$.
Among these, nine were randomly selected for re-simulation.

\subsection{Replacement by galaxy model} \label{gal_model}

In this subsection, we describe how the dry merger process is simulated in the cluster regions using galaxy models.
In these simulations, the particles within twice the virial radius of each cluster halo were traced back to $z=2.85$.
The virialized dark matter haloes at $z=2.85$ for the traced particles were identified by the friends-of-friends (FOF) algorithm (\citealt{1985ApJ...292..371D}).
For the dark matter haloes with FOF halo masses exceeding $10^{11}\mathrm{M}_{\odot}$,
the spherical regions within the virial radius from the halo centres of mass were replaced with our galaxy model, as described below.

In this replacement, we do not take into account the haloes which are already the 
substructures of the larger, FOF haloes at $z=2.85$.
Since these substructures could host galaxies, these may affect the galaxy merger rate.
To investigate these effects, we counted the substructures with mass $\geq 10^{12}~M_{\odot}$,
which could host galaxies with mass $\ga 10^{10}~M_{\odot}$ at $z=2.85$.
To do this, we use the Spline Kernel Interpolative DENMAX (SKID\footnote[1]{http://www-hpcc.astro.washington.edu/tools/skid.html}) algorithm (\citealt{1994ApJ...436..467G}; \citealt{2005MNRAS.363....2K}) (see also \S 2.4).
We found that the ratio of these substructures to the total haloes we replace with our galaxy model is less than $0.5$.
This indicates that these substructures do not affect significantly the galaxy merger rate.

Our model galaxies consisted of two components, i.e. a stellar bulge and a dark matter halo.
The mass distributions in each component were modeled by a two-component Hernquist profile (\citealt{1996ApJ...471...68C}).
A similar model was used in paper I, however, different bulge mass assumptions were imposed on the galaxy model.
In the present paper, the bulge mass $M_{*}$ and the dark matter halo $M_{\mathrm{halo}}$ were based on the stellar-to-mass relation obtained by the abundance matching method of \citet{2010ApJ...710..903M}, assuming that $M_{\mathrm{halo}}$ and $M_{*}$ sum to $M_{\mathrm{vir}}$.

\subsubsection{Dark matter halo}
The dark matter haloes were modeled by the Navarro--Frenk--White (NFW) profile obtained in cosmological simulations \citep*{1997ApJ...490..493N}.
The concentration parameter $c_{\mathrm{vir}}$ was derived using the approach of \citet{2009ApJ...696.1094R}, which is a modified version of a toy model proposed by \citet{2001MNRAS.321..559B} that sets $c_{\mathrm{vir}} = c_{\mathrm{vir}} (M_{\mathrm{vir}}, z)$.

The dark matter mass distribution was expressed as a \citet{1990ApJ...356..359H} profile of the total mass, $M_{\mathrm{halo}}$:
\begin{equation}
  \label{eq:hern_dm}
  \rho_{\mathrm{dm}} (r) = \frac {M_{\mathrm{halo}}} {2\pi} \frac {a_{\mathrm{dm}}} {r(r+a_{\mathrm{dm}})^3},
\end{equation}
scaled to match the abovementioned NFW model.
The cumulative mass profile corresponding to Equation (\ref{eq:hern_dm}) is
\begin{equation}
  M_{\mathrm{dm}} (<r) = M_{\mathrm{halo}} \frac{r^2} {(r+a_{\mathrm{dm}})^2}.
\end{equation}
In this model, the total mass within the virial radius of a dark matter halo, $M_{\mathrm{dm}} (<r_{\mathrm{vir}})$, corresponds to $M_{\mathrm{halo}}$ (the mass within the virial radius of the NFW model).
The inner density profile of the Hernquist profile also matches that of the NFW model.
To Satisfy the abovementioned condition, the scale radius of the Hernquist profile $a_{\mathrm{dm}}$ was expressed as the following function of $c_{\mathrm{vir}}$ and the scalelength $r_{\mathrm{s}}$ ($r_{\mathrm{s}}=r_{\mathrm{vir}}/c_{\mathrm{vir}}$) of the NFW halo:
\begin{equation}
  a_{\mathrm{dm}} = \left( \frac{1} {r_{\mathrm{s}} \sqrt{2[\ln(1+c_{\mathrm{vir}})-c_{\mathrm{vir}}/(1+c_{\mathrm{vir}})]}}  - \frac{1} {r_{\mathrm{vir}}}    \right)^{-1}.
\end{equation}
The parameters $c_{\mathrm{vir}}$ and $r_{\mathrm{s}}$ were given by our NFW model, whereas $r_{\mathrm{vir}}$ was obtained by our cosmological simulation results.
Thus, we could derive $a_{\mathrm{dm}}$ and obtain the corresponding halo model for our galaxy model.

\subsubsection{Stellar Bulge}

The initial mass distribution of the bulge was modeled by a Hernquist profile:
\begin{equation}
  \label{eq:hern_star}
  \rho_{*} (r) = \frac {M_{*}} {2\pi} \frac {a_*} {r(r+a_*)^3}.
\end{equation}
In Equation (\ref{eq:hern_star}), $M_{*}$ and $a_*$ denote the total stellar mass and scale radius, respectively.
The effective radius $R_\mathrm{e}$ is taken as the projected radius enclosing half of the stellar mass.
$R_\mathrm{e}$ is related to $a_*$ by $R_e = 1.8153a_{*}$.
For the effective radius, we adopted the observed stellar mass--size relation of ETGs at $2.0 < z < 2.5$ (\citealt{2012ApJ...746..162N}):
\begin{equation}
  \log R_\mathrm{e} = 0.04 + 0.69(\log M_* -11).
\end{equation}
In the bulge, we computed the abovementioned relation in terms of the mean size, ignoring the scatter for simplicity.
The validity of this assumption is open to debate.
\citet{2006ApJ...649L..71K} found that $\sim45$ per cent of massive galaxies at $z\sim2.3$ have quiescent star formation 
activity (see also \citealt{2008ApJ...677..219K}).
More recently, \citet{2013ApJ...777...18M} showed that the passive fraction is about 30 per cent at the highest mass end of the luminosity function at $z=2$.
Thus, our simulated galaxies actually could be star forming galaxies even for the most massive galaxies.
In general, star forming galaxies have lager sizes than passive galaxies at $z\sim2$ (\citealt{2014ApJ...788...28V}).
Therefore, we may underestimate the sizes of simulated galaxies.
Here, we neglect these possibilities.

Finally, we ensured that the bulge and the dark matter halo were equilibrated in the total gravitational potential by the $N$-body method.
For further details of this method, see paper I and \citet{1994MNRAS.269...13K}.

\begin{figure}
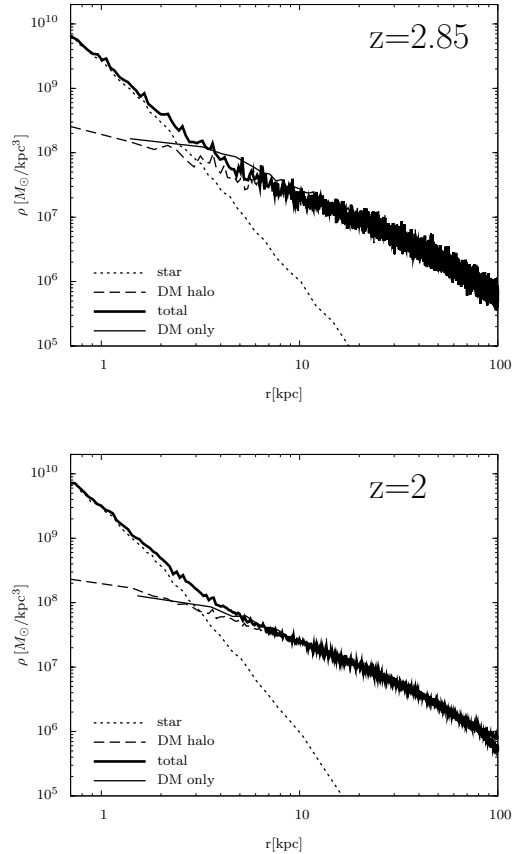

  \begin{minipage}{84mm}
    \begin{center}
      \begin{tabular}{c}
        \hspace{-15mm}
        \resizebox{84mm}{!}{\input{./picture/profile_comparison_z2p85_H07.tex}}  \\
        \hspace{-15mm}
        \resizebox{84mm}{!}{\input{./picture/profile_comparison_z2_H07.tex}}  
      \end{tabular}
      \caption{Density profiles of the central galaxy for the run H07 at $z=2$ and 2.85.
      The thick solid lines are the density profiles of all stars and all dark matter particles.
      The dotted (dashed) lines denote star (dark matter) particles.
      The density profiles of dark matter in the dark matter only simulation are also plotted (the thin solid lines).
      }
      \label{profile_test_H07}
    \end{center}
  \end{minipage}
\end{figure}

\begin{figure}
  \begin{minipage}{84mm}
    \begin{center}
      \begin{tabular}{c}
        \hspace{-15mm}
        \resizebox{84mm}{!}{\input{./picture/profile_comparison_z2p85_H29.tex}}  \\
        \hspace{-15mm}
        \resizebox{84mm}{!}{\input{./picture/profile_comparison_z2_H29.tex}}
      \end{tabular}
      \caption{The same as Figure \ref{profile_test_H07}, but for H29.
      }
      \label{profile_test_H29}
    \end{center}
  \end{minipage}
\end{figure}

\subsection{Re-simulation} \label{re_sim}
Using our galaxy models, we re-simulated the cluster regions extracted from our cosmological simulation.
After replacing the virialized dark matter haloes in our cosmological simulation with the  higher resolution galaxy models at $z=2.85$, we simulated the cluster regions from $z=2.85$ to $z=0$.
In the galaxy models, the resolution of the star and dark matter particles was improved by a factor of 8 compared to that of the dark matter particles in the cosmological $N$-body simulation, and corresponds to a mass resolution of $6.25\times 10^6 \mathrm{M}_{\odot}$.
The higher resolution dark matter and star particles of the galaxy models have the same mass in order to minimize differential two-body heating effects between the two components.
The softening length of both the particles was 0.325 kpc.
The particle mass and softening length are comparable with  those of the Phoenix simulations (\citealt{2012MNRAS.425.2169G}; \citealt{2013MNRAS.435..901L}).
The remaining dark matter particles beyond the virialized haloes were retained with a mass of $5.0\times10^7\mathrm{M}_{\odot}$, and the softening length was 0.65 kpc.
These softening lengths were constant in the physical coordinates from $z=2.85$ to $z=0$.
Simulations were conducted in GADGET-2 code (\citealt{2005MNRAS.364.1105S}), a parallelized TreePM code.
To mimic the tidal fields, we re-simulated the Lagrangian region of the particles residing within twice the virial radius of the cluster haloes at $z=0$.
Vacuum boundary conditions were assumed in the Lagrangian region.
To confirm that the vacuum boundary condition does not affect our main results, we modeled the dark matter distributions in cluster haloes by a cosmological simulation with periodic boundary conditions, and compared the results with our re-simulated results.

As a test, we investigated  the evolution of our galaxy models from $z=2.85$ to $z=2$ after the replacement.
We compare them with the dark matter haloes in the dark matter only simulations in Fig. \ref{profile_test_H07} and \ref{profile_test_H29}.
We have confirmed that our galaxy models are stable and that the total density profiles do not differ significantly 
at $r \ga 30$kpc between our galaxy models and the dark matter haloes in the dark matter only cases.
The latter result shows that our procedure is valid in this paper.

\begin{figure*}
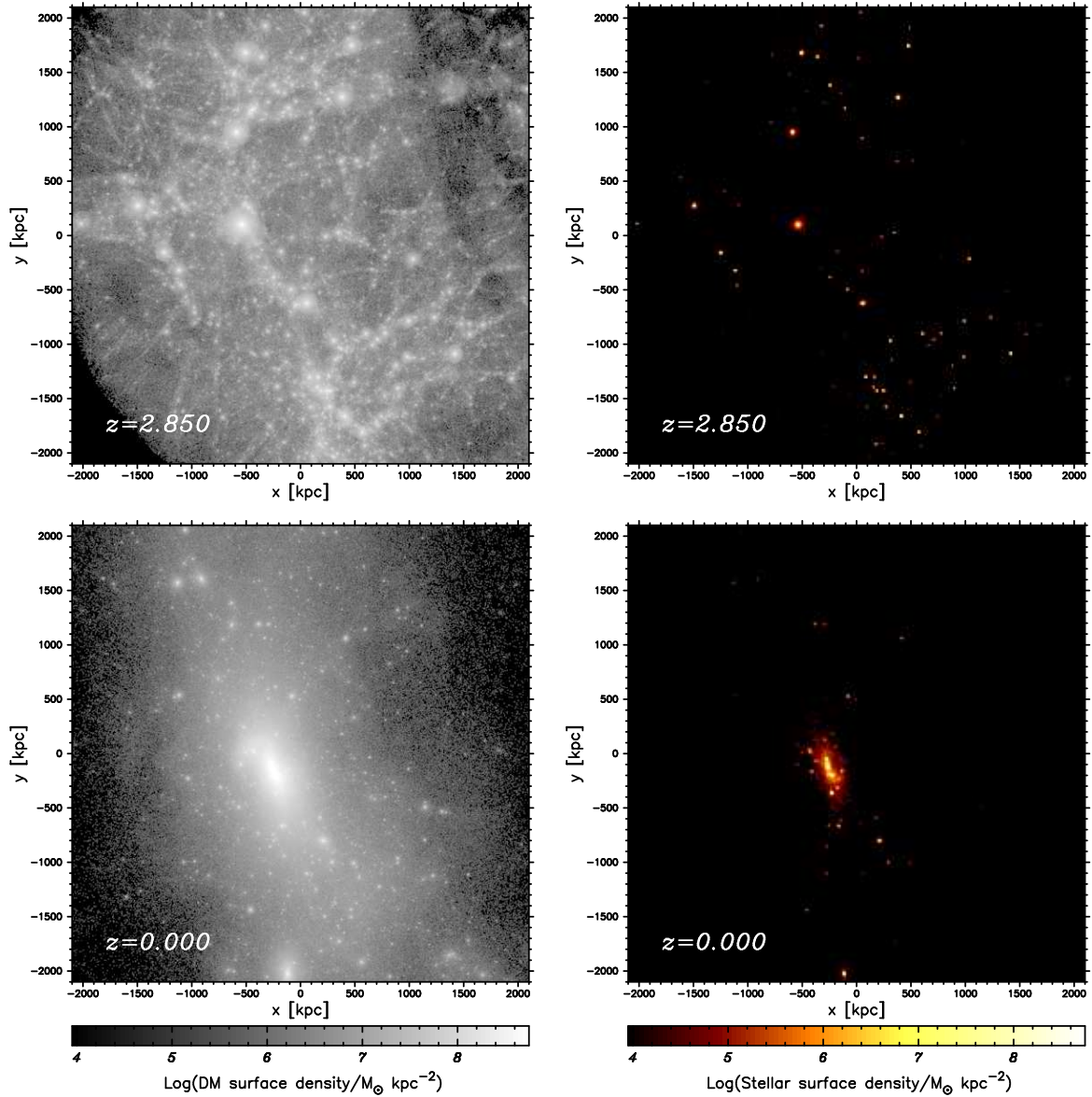

    \begin{minipage}{180mm}
        \begin{center}
            \begin{tabular}{cc}
                \hspace{-8mm}
                \resizebox{75mm}{!}{\includegraphics[angle=-90]{./picture/snap000_dm_H29_kpc.eps}} &
                \resizebox{75mm}{!}{\includegraphics[angle=-90]{./picture/snap000_star_H29_kpc.eps}} \\
                \hspace{-8mm}
                \resizebox{75mm}{!}{\includegraphics[angle=-90]{./picture/snap037_dm_H29_kpc.eps}} &
                \resizebox{75mm}{!}{\includegraphics[angle=-90]{./picture/snap037_star_H29_kpc.eps}}
            \end{tabular}
            \caption{
            Surface density maps of the dark matter particles (left) and star particles (right) at $z=2.85$ (top) and $z=0$ (bottom) in run H29 (see text). The star particles reside in dark haloes with FOF halo mass $M_\mathrm{FOF} \ge 10^{11} \mathrm{M}_{\odot}$ at $z=2.85$ as described in the text. The box size is 4 Mpc $\times $ 4  Mpc on the physical scale.  Gray scale and color bars indicate the surface densities of dark matter and stellar matter, respectively.      
            (A color version of this figure is available in the  online journal.)
            }
            \label{snapshots}
        \end{center}
    \end{minipage}
\end{figure*}

\begin{table*}
  \centering
  \begin{minipage}[t]{180mm}
    \caption{
    Simulated physical properties of cluster haloes at $z=0$, central haloes (see text) at $z=2.85$ and the central galaxies in the cluster haloes at $z=0$ and $z=2.85$ (see text).
    The masses and sizes of the stellar components in the central galaxies are based on $r_{\rm trunc} = $ 60 kpc.
    $M_{\mathrm{halo}}$ and $M_{*}$ are the masses of the dark matter halo and the stellar systems of central galaxies, respectively.
    $\alpha$ is the size growth efficiency defined in the text.
    The `avg.' row gives the averaged values (values in parentheses are averaged without H48 and H49).
     }
    \label{table:size_growth_efficiency}
    \begin{tabular}{@{}cccccccccccc@{}}
      \hline
      \hline
      Name & $M_{\mathrm{halo}}(z=2.85)$ & $M_{\mathrm{halo}} $& $M_{*} (z=2.85)$ & $M_{*} (z=0)$ & $\frac{M_{*} (z=0)}  {M_{*} (z=2)}$ & $\frac{R_{e} (z=0)} {R_{e} (z=2)}$ &  $\alpha$  \\
       & ($10^{13} \mathrm{M}_{\odot}$)& ($10^{14} \mathrm{M}_{\odot}$)  & ($10^{10} \mathrm{M}_{\odot}$) & ($10^{10} \mathrm{M}_{\odot}$) &  &  &  $(R_e \propto M_{*}^{\alpha})$   \\

      \hline
      H07   & 1.28 & $3.73$ & 10.7 & 22.2 & 2.12 & 2.56 & 1.25  \\
      H29   & 1.03 & $1.81$ & 9.37 & 15.1 & 1.66 & 2.41 & 1.74  \\
      H31   & 0.594 & $1.57$ & 6.44 & 17.1 & 2.71 & 6.34 & 1.85  \\
      H33   & 0.320 & $1.22$ & 3.61 & 11.2 & 3.17 & 5.46 & 1.47  \\
      H43   & 0.797 & $1.14$ & 7.80 & 17.1 & 2.24 & 3.77 & 1.64  \\
      H44   & 0.365 & $0.826$ & 4.12 & 9.48 & 2.35 & 4.68 & 1.81 \\
      H45   & 0.398 & $1.17$ & 4.58 & 15.7 & 3.51 & 5.84 & 1.41  \\
      H48   & 1.34 & $1.10$ & 11.0 & 11.4 & 1.06 & 1.31 & 4.69  \\
      H49   & 0.845 & $0.879$ & 8.19 & 8.43 & 1.05 & 1.25 & 4.32  \\
      \hline
      avg.   &     -          & -  &   -    &     -    & 2.27(2.59) & 3.98 (4.70)& 1.90 (1.63) \\
      \hline
      \hline
    \end{tabular}
  \end{minipage}
\end{table*}

\subsection{Identification of galaxies and galaxy mergers} \label{merger_model}
Galaxies and their merger histories were identified from 38 simulation snapshots collected from $z=2.85$ to $z=0$.
The time interval between the snapshots was $\sim$100 Myrs at $z=2.85$ and  $\sim$500 Myrs at $z=0$.
Galaxies were identified using the SKID algorithm, which evaluates the particle number densities at each particle position, and moves each particle along the particle number density gradient until it oscillates around some positions.
The particles are then linked using the FOF algorithm.
Finally, in each linked group, particles with negative energy binding to the group centre of mass are retained; those that violate this criterion are removed.
This procedure is applied to all star particles.
Groups with more than 100 particles are defined as galaxies.
As described in paper I, galaxies are surrounded by sparse stellar particles, which may affect their half-mass radii and effective radii.
Thus, imitating paper I, the galactic stellar mass and radii were defined by imposing a truncation radius $r_{\mathrm{trunc}} = 60$ kpc (see paper I for details).
We define the galaxy stellar mass as the stellar mass within $r_{\mathrm{trunc}}$, which is centred at the galaxy's centre of mass.
To avoid the effects of substructures on the galaxy mass and size, we get rid of the substructures in this analysis.

In constructing merger trees of galaxies, we connected a galaxy to its descendant galaxy at a later redshift through simulation snapshots.
The most tightly bound particles in a galaxy was identified at a given redshift and traced to the next snapshot.
The galaxy containing that particle in the next snapshot was assumed to be the descendant galaxy.
If the most bound particles in multiple galaxies were found in a single descendant galaxy, the earlier galaxies were assumed to have merged into the descendant galaxy.
We call the most massive galaxy in each cluster halo at $z=0$ the {\it central galaxy}.
When tracing the merger history of the central galaxy through earlier snapshots, the most massive progenitor at each snapshot (or redshift) was similarly called the central galaxy.
Furthermore, the halo containing the central galaxy is defined as the {\it central halo}.

\section{Results}

To investigate the effect of dry mergers on the mass and size growth of ETGs in dense environments, we re-simulate nine of the generated cluster haloes.
Their virial masses were $M_{\mathrm{vir}} \simeq 10^{14} - 5\times 10^{14} \mathrm{M}_{\odot}$ at $z=0$.
Their masses, physical properties of their most massive galaxies at $z=0$ and their most massive progenitors (hereafter called central galaxies; see \S \ref{merger_model}) are summarized in Table \ref{table:size_growth_efficiency}.
The dark matter distribution and star particles in a typical cluster halo (run H29) at $z=2.85$ and $z=0$ are presented in Fig. \ref{snapshots}.
Most of the star particles initialized at $z=2.85$ reside in the central region of the cluster halo at $z=0$, and are distributed along the elongation direction of the cluster halo.
The stellar mass of the central galaxy is $1.96 \times 10^{11} \mathrm{M}_{\odot}$ at $z=0$.
There are many dark matter subhaloes in the cluster haloes at $z=0$ (Fig. \ref{snapshots}; bottom-left).
While there are numerous star particles in subhaloes near the centre of the cluster halo, they are sparse in more distant subhaloes.
This spatial difference in the number ratio of star particles to dark matter particles indicates that old stellar populations mainly reside in the high density regions of galaxies.
This distribution should be related to the morphology--density relation observed in the local Universe (e.g. \citealt{1980ApJ...236..351D}).

To investigate whether the galaxy sizes are stable, we checked the change of the 
effective radius of each central galaxy from $z=2.85$ to the time of the first 
interaction with another galaxy.
We found that the satellite galaxies with stellar masses $M_{*} \la 4\times 10^{10} M_{\odot}$ 
increase their sizes up to $\sim0.8$kpc artificially because of insufficient mass and spatial resolutions.
For central galaxies, these changes are smaller than those due to mergers in all runs.
Thus, in the following subsections, we focus on the evolution of the central galaxies from $z=2$ to $z=0$.

\subsection{Mass and size growth of the central  galaxies}

Table \ref{table:size_growth_efficiency} lists the mass growth factors $M_{*} (z=0)/M_{*} (z=2)$ of the central galaxies.
The diversity of the mass growth factors reflects their various merger histories.
Most of the central galaxies increased their masses by a factor of 2-3 from $z=2$ to $z=0$ (the average stellar mass growth factor is 2.27), which is consistent with the observational estimates of \citet{2010ApJ...709.1018V}.
They selected sample galaxies with constant number density in the redshift range $0\la z \la 2$ from the NOAO/Yale NEWFIRM Medium Band Survey, and estimated the mass increase of the compact massive ETGs from $z=2$ to $z=0$.
Our simulation naturally reproduces their estimated value.

Table \ref{table:size_growth_efficiency} also shows the size growth factors $R_{e} (z=0)/R_{e} (z=2)$.
The average size growth factor of the central galaxies is 3.98 from $z=2$ to $z=0$, which is also consistent with the observational estimates of \citet{2010ApJ...709.1018V}.
This growth can be approximated as $R_e\propto (1+z)^{-a}$ with $a \sim 1.26$.
In the observational estimates, $a$ ranges from 0.75 (\citealt{2010ApJ...717L.103N}) to 1.62 (\citealt{2011ApJ...739L..44D}, see also \citealt{2008ApJ...687L..61B}; \citealt{2012ApJ...746..162N}; \citealt*{2012MNRAS.422L..62C}; \citealt{2014ApJ...788...28V}).
In paper I, we reported that changes in $R_e$ reflect the manner of the mass deposition of the accreted stars.
As we will show in \S \ref{sec:density_profile}, stars accreted through the mergers are frequently, deposited in the outer regions of the central galaxies.
This process ensures efficient size growth.

The efficiency $\alpha$ of the size growth is important for evaluating the size evolution of ETGs.
The radial and mass growths are related as follows:
\begin{equation}
  \label{eq:efficiency}
  \frac {R_{e,f}} {R_{e,i}} = \left( \frac {M_{*, f}} {M_{*,i}} \right)^{\alpha},
\end{equation}
where, $R_{e,i}$ and $R_{e,f}$ are the effective radius at $z=2$ and $z=0$, $M_{*, i}$ and $M_{*,f}$ are the stellar mass at at $z=2$ and $z=0$, and  $\alpha$ is derived from our numerical results, as in paper I.
The size growth efficiencies $\alpha$ of the central galaxies are listed in Table \ref{table:size_growth_efficiency}.
The average size growth efficiency $<\alpha> \simeq 1.9$ is in good agreement with the observational estimates of \citet{2010ApJ...709.1018V}, and is roughly consistent with the observational constraint given by \citet{2009ApJ...697.1290B}.
In addition, our $<\alpha>$ also agrees with those derived from previous numerical experiments in a cosmological context (\citealt{2009ApJ...699L.178N}; \citealt{2012ApJ...744...63O}; \citealt{2012MNRAS.424..747L}).
In paper I, the maximum efficiency was $\alpha \simeq 2.7$ obtained for sequential dry minor mergers of compact satellites.
However, paper I assumed the typical mass ratios of minor mergers ($1/20 < M_2 / M_1 \leq 10$) and radially orbiting satellites.
In our present simulations, the mass ratios of the mergers vary, and the merger orbits are obtained from the numerical results of the cosmological simulations at $z=2.85$.
In \S \ref{sec:merger_histories}, we will demonstrate that the mass and size growth of our simulated central galaxies is mainly contributed by dry major mergers.
This may explain why the size growth efficiency $<\alpha> \simeq 1.9$ is lower in the present paper than that in paper I, which was limited to dry minor mergers.

As discussed above, the average mass growth, size growth and size growth efficiency derived in this paper agree with observational estimations.
This indicates that dry mergers sufficiently explain the size growth of compact massive ETGs.
Because the mass and size growths of central galaxies widely vary, they are related to the stellar merger histories of the galaxies in the next subsection.

\subsection{Stellar merger histories}\label{sec:merger_histories}
\begin{table}
  \centering
  \begin{minipage}[t]{80mm}
    \caption{Number of major mergers and the mass fractions  of  major and minor dry mergers in our simulated central galaxies at $z=0$.
    Columns 2--5 list the number of merger events with mass ratios $M_{2}/M_{1} \simeq 1$, $1/2$, $1/3$ and $1/4$, respectively.
    Also listed are the mass fractions of dry major mergers with mass ratios $1/4 < M_{2}/M_{1} < 1$ (Column 6) and of dry minor mergers with the mass ratios $M_{2}/M_{1} < 1/4$ (Column 7).
    }
    \label{table:merger_mass_ratio}
    \begin{tabular}{@{}cccccccccccc@{}}
      \hline
      \hline
      Name & 1 & 1/2 & 1/3 & 1/4 & $f_{\mathrm{major}}$ &  $f_{\mathrm{minor}}$  
      & $\frac{M_*(z=2.85)}{M_*(z=0)}$\\

      \hline
      H07   & 1 & 0 & 0 & 0 & 0.46 & 0.053  & 0.48\\
      H29   & 0 & 1 & 0 & 0 & 0.29 & 0.089  & 0.62\\
      H31   & 0 & 1 & 0 & 1 & 0.55 & 0.070  & 0.38\\
      H33   & 2 & 0 & 0 & 0 & 0.63 & 0.046  &0.32\\
      H43   & 0 & 2 & 0 & 0 & 0.54 & 0  &0.46\\
      H44   & 0 & 1 & 1 & 0 & 0.55 & 0.011  &0.43\\
      H45   & 0 & 1 & 1 & 0 & 0.72 & 0  &0.29\\
      H48   & 0 & 0 & 0 & 0 & 0 & 0.032  & 0.96\\
      H49   & 0 & 0 & 0 & 0 & 0 & 0.028  & 0.97\\
      \hline
      \hline
    \end{tabular}
  \end{minipage}
\end{table}

\begin{figure}
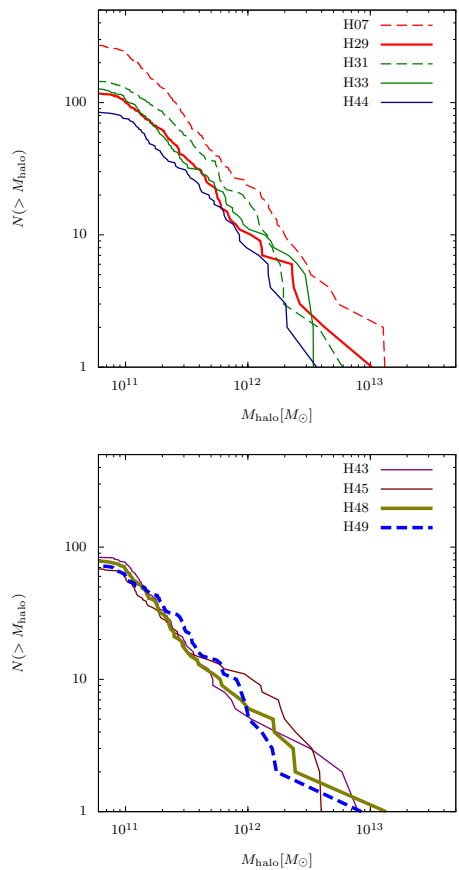

  \resizebox{84mm}{!}{\input{./picture/m_halo_dist_2_2.tex}}  \\
  \resizebox{84mm}{!}{\input{./picture/m_halo_dist_1_2.tex}}  
  \caption{Cumulative numbers of dark matter haloes  as functions of their mass at $z=2.85$ in each run. 
  (A color version of this figure is available in the  online journal.)
  }
  \label{z_mass_size_all_central}
\end{figure}

We now analyse the stellar merger histories of the central galaxies, identifying  the merger events and their mass ratios from our simulation results.
Mass growth of the central galaxies occurs mainly by mergers with mass ratios $M_{2}/M_{1} \geq 1/4$, where $M_1$ and $M_2$ are the masses of the mergers.
Table \ref{table:merger_mass_ratio} tabulates the number of merger events at different mass ratios;
$1\geq M_{2}/M_{1} > 1/1.5$ (column `1'), $1/1.5\geq M_{2}/M_{1} > 1/2.5$ (column `1/2'), $1/2.5 \geq M_{2}/M_{1} > 1/3.5$ (column `1/3') and $1/3.5\geq M_{2}/M_{1} \geq 1/4$ (column `1/4').
As in previous studies, mergers with mass ratios $M_{2}/M_{1} \geq 1/4$ are classified as major mergers and those with smaller mass ratios are called minor mergers (e.g. paper I and references therein).
The central galaxies (except H48 and H49) experienced one or two dry major mergers.
For comparison, the cumulative mass functions of the virialized haloes at $z=2.85$ are plotted in Fig. \ref{z_mass_size_all_central}.
There are roughly one hundred virialized haloes in each run at $z=2.85$.
The cumulative mass functions can be approximated as
\begin{equation}
  N(>M_{\mathrm{halo}}) \propto M_{\mathrm{halo}}^{-\beta}
\end{equation}
with $\beta \sim 1.0$ -- $1.2$
, and few of the virialized dark haloes exceed 1/4 times the mass of the dark haloes of the central galaxies.
In run H48 and H49, the mass of the second massive dark halo is roughly one-fifth smaller than the first massive dark halo at $z=2.85$.
Consequently, there is no dry major merger in these cluster haloes.

The mass fraction accreted by dry mergers in the central galaxies at $z=0$, denoted as $f_{\mathrm{major}}$ is given by
\begin{equation}
  \label{eq:mass_fraction_minor}
  f_{\mathrm{major}} = \frac {\Delta M_{\mathrm{major}}} {M_{*}(z=0)},
\end{equation}
where $\Delta M_{\mathrm{major}}$ is the mass introduced by the major mergers from $z=2.85$ to $z=0$.
$f_{\mathrm{minor}}$ is given by 
\begin{equation}
  \label{eq:mass_fraction_minor}
  f_{\mathrm{minor}} = \frac {M_{*}(z=0) - M_{*}(z=2.85)-\Delta M_{\mathrm{major}}} {M_{*}(z=0)}.
\end{equation}
In most central galaxies $f_{\mathrm{major}}$ is larger than $f_{\mathrm{minor}}$ (see Table \ref{table:merger_mass_ratio}).
In \S \ref{sec:mass_size_relation}, we will show that major mergers mainly occur at $z<1$.
According to these results, the mass and size of central galaxies in dense environments usually grows by dry major mergers.
The exceptions are H48 and H49, in which $f_{\mathrm{major}}=0$ and the stellar mass increase of the central galaxies is small from $z=2.85$ onward.
In these galaxies, mass accretes by minor mergers (see Table \ref{table:merger_mass_ratio}).

\subsection{Density profiles}\label{sec:density_profile}

\begin{figure*}
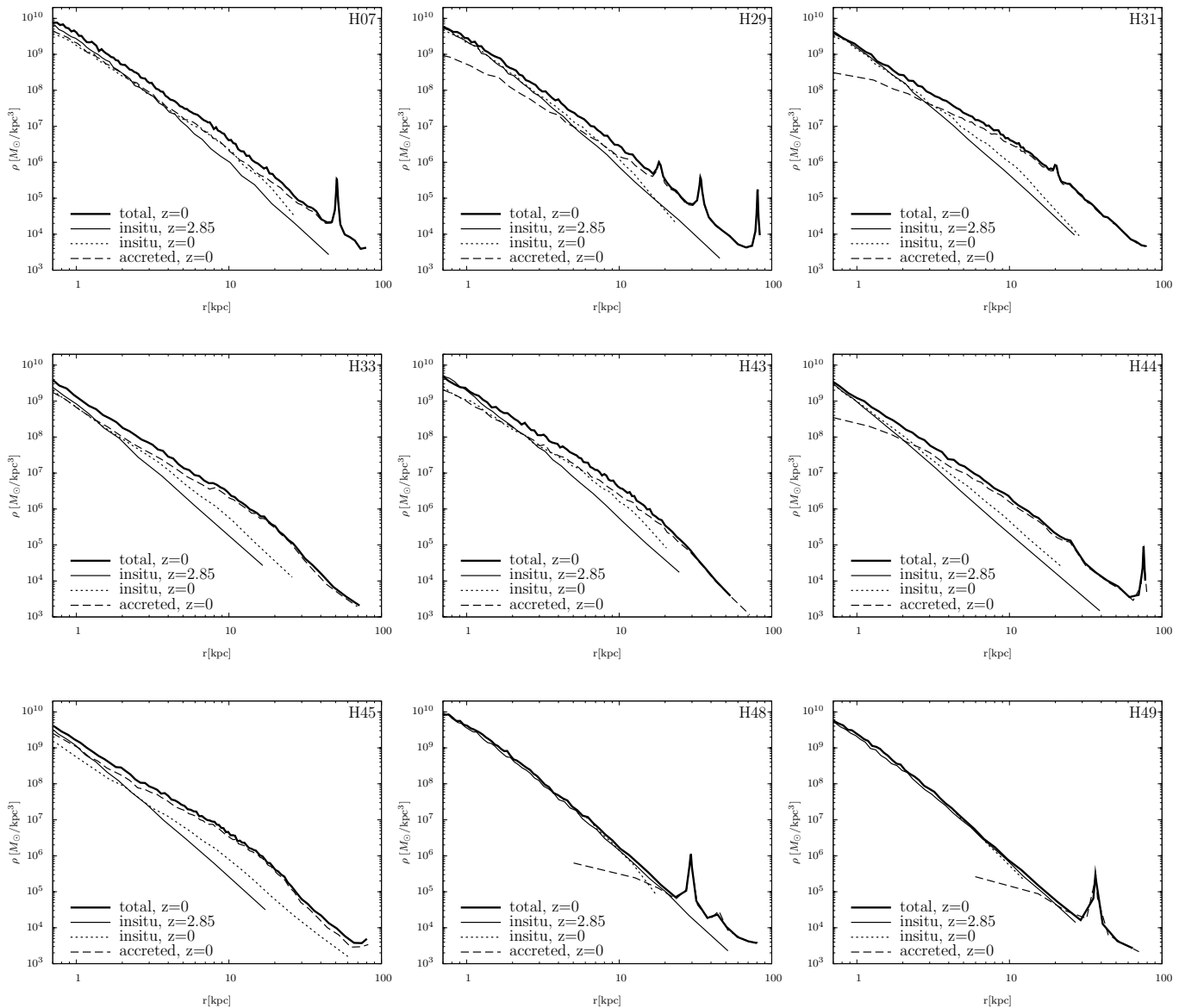

    \begin{minipage}{180mm}
        \begin{center}
            \begin{tabular}{ccc}
                \hspace{-17mm}
                \resizebox{75mm}{!}{\input{./picture/r_log_rho_log_H07_10.tex}} &
                \hspace{-20mm}
                \resizebox{75mm}{!}{\input{./picture/r_log_rho_log_H29_08.tex}} &
                \hspace{-20mm}
                \resizebox{75mm}{!}{\input{./picture/r_log_rho_log_H31_03.tex}} \\
                \hspace{-17mm}
                \resizebox{75mm}{!}{\input{./picture/r_log_rho_log_H33_04.tex}} &
                \hspace{-20mm}
                \resizebox{75mm}{!}{\input{./picture/r_log_rho_log_H43_01.tex}} &
                \hspace{-20mm}
                \resizebox{75mm}{!}{\input{./picture/r_log_rho_log_H44_01.tex}} \\
                \hspace{-17mm}
                \resizebox{75mm}{!}{\input{./picture/r_log_rho_log_H45_02.tex}} &
                \hspace{-20mm}
                \resizebox{75mm}{!}{\input{./picture/r_log_rho_log_H48_04.tex}} &
                \hspace{-20mm}
                \resizebox{75mm}{!}{\input{./picture/r_log_rho_log_H49_05.tex}} \\
            \end{tabular}
            \caption{
            Stellar density profiles of the central galaxies at $z=0$ in the cluster haloes.
            From top left to bottom right: runs H07, H29, H31, H33, H43, H44, H45, H48 and H49.
            Thick solid lines are the density profiles of all stars in the central galaxies, dotted and dashed lines denote stars originated in the most massive progenitor (insitu stars) and the less massive galaxies (accreted stars), respectively.
            Thin lines show the initial profiles at $z=2.85$.
            }
            \label{density_profiles}
        \end{center}
    \end{minipage}
\end{figure*}

\begin{table}
  \centering
  \begin{minipage}[t]{80mm}
    \caption{Half mass radii of insitu stars and accreted stars, the concentrations, $\mathrm{C}=r_{\mathrm{trunc}}/r_{\mathrm{half}}$, of insitu stars and accreted stars, and concentration ratios, $R=\mathrm{C_{insitu}} / \mathrm{C_{accreted}}$, of the central galaxies at $z=0$.
    Here, $r_{\mathrm{trunc}}=60$kpc (see text).
    }
    \label{table:concentration_ratio}
    \begin{tabular}{@{}cccccc@{}}
      \hline
      \hline
      Name & $r_{\mathrm{half, insitu}}$ & $r_{\mathrm{half, accreted}}$ & $\mathrm{C_{insitu}}$ & $\mathrm{C_{accreted}}$ & $R$\\
      & (kpc) & (kpc) & & &  \\

      \hline
      H07   & 3.61 & 4.87 & 16.6 & 12.3 & 1.35\\
      H29   & 2.13 & 9.00 & 28.1 & 6.67 & 4.21\\
      H31   & 2.35 & 13.1 & 25.5 & 4.57 & 5.58\\
      H33   & 7.64 & 6.03 & 7.85 & 9.96 & 0.788\\
      H43   & 4.03 & 7.01 & 14.9 & 8.55 & 1.74\\
      H44   & 1.78 & 9.31 & 33.7 & 6.44 & 5.23\\
      H45   & 4.49 & 8.26 & 13.3 & 7.27 & 1.83\\
      H48   & 1.85 & 25.4 & 32.4 & 2.36 & 13.7\\
      H49   & 1.61 & 25.4 & 37.2 & 2.36 & 15.8\\
      \hline
      \hline
    \end{tabular}
  \end{minipage}
\end{table}

The time evolution of the stellar density profiles is very important for understanding the size growth of the central galaxies.
The thick lines in Fig. \ref{density_profiles} show the angle-averaged stellar density profiles of the central galaxies at $z=0$ for nine runs (H07, H29, H31, H33, H43, H44, H45, H48 and H49).
Three individual stellar density profiles are also plotted;
the initial density profiles of the central galaxies at $z=2.85$ (called {\it insitu stars}), the density profiles of stars at $z=0$ that resided in the central galaxies at $z=2.85$ (also called insitu stars) and the density profiles of stars at $z=0$ that resided in other virialized haloes at $z=2.85$ (called {\it accreted stars}).

As shown in these figures, the stellar density profiles can be grouped into three types.
In the first type, the stellar density profile shows excesses at $r > 1$ kpc from the progenitor stellar density profile at $z=2.85$, but shows little change in the density profile at $r < 1$ kpc.
The stellar density profiles of H29, H31, H43 and H44 are of this type.
The size growth efficiencies of these runs are $\alpha =$ 1.71--1.89 (see Table \ref{table:size_growth_efficiency}).
The accreted stars dominate in the excesses of the stellar density profiles at $r > 1$ kpc, whereas the insitu stars dominate in the inner part of the central galaxies (except for H43).
The radii of the regions dominated by insitu stars range from 2 to 5 kpc, as shown in Fig. \ref{density_profiles}.
These profiles well agree with the merger remnants yielded  by $N$-body simulations of dry minor mergers (paper I; \citealt*{2013MNRAS.429.2924H}; \citealt{2013MNRAS.431..767B}).
In H43, the insitu and accreted stars make comparable contributions to the stellar density profile at $z=0$ over a wide region ($r<6 $ kpc).
This feature typifies equal mass major merger remnants (see Appendix A).
As  shown in \S \ref{sec:merger_histories}, the merger mass ratios of type 1 profiles are $1/4 < M_{2}/M_{1} < 1/2$, signifying major mergers.

In the second type, stellar density profile excesses appear at all radii from the progenitor stellar density profile.
H07, H33 and H45 are of this type.
The size growth efficiencies of these runs are $\alpha =$ 1.29--1.49 (see Table \ref{table:size_growth_efficiency}).
The accreted and insitu stars contribute similarly to the stellar densities at $r < 1$ kpc, which is typical of equal mass major merger remnants.
The central galaxies in H07 and H33 experienced one or two equal mass mergers (see Table \ref{table:merger_mass_ratio}).
In H45, the accreted stars dominate the stellar density profile at all radii because the central galaxy experienced multiple major mergers, and the accreted stars of two companion galaxies considerably contribute to the stellar density of the remnant.
Consequently, the mass fraction of major mergers is large in this case ($f_{\mathrm{major}} = 0.71$; Table \ref{table:merger_mass_ratio}).

In the third type of density profile, the accreted stars make a minor contribution, manifesting mainly as spikes in the outer galaxy profiles.
The central galaxies have experienced only a few minor mergers from $z=2.85$ to $z=0$ (see Table \ref{table:merger_mass_ratio}).
Runs H48 and H49 show type 3 density profiles.
Because of the minor mergers, their size growth efficiencies ($\alpha =$ 2.84 and 2.78 in runs H48 and H49, respectively) are higher than those in other runs.
The spikes in the outer profile represent the surviving cores of satellite galaxies.
These cores might artificially elevate the growth efficiencies in our analysis.
However, we have already rejected this possibility in paper I, by evaluating the effect of such cores on the growth efficiency.
Although the cluster halo masses of H48 and H49 are comparable to those of H44 and H45, their central galaxies remain compact, which also occurs in high-z ETGs.
These compact galaxies resemble compact ETGs found in the local Universe (\citealt{2013ApJ...762...77P}).
However, our compact galaxies could reflect our initial condition about the sizes of galaxies.
We assume the initial galaxy sizes as the median sizes of the stellar mass-size relation of ETGs given by \citet{2012ApJ...746..162N}.
Thus, the initial galaxy sizes in our simulations are compact.
However, \citet{2014ApJ...788...28V} find the intrinsic scatter in size for both early- and late-type galaxies 
over the redshift range $0<z<3$.
Furthermore, \citet{2013ApJ...777...18M} showed that the passive fraction is about 30 per cent 
at the highest mass end of the luminosity function at z=2 (see also \S \ref{gal_model}).
Thus, we do not conclude that our compact ETGs at z=0 are analogues of those observed 
by \citet{2013ApJ...762...77P}.

These compact galaxies resemble compact ETGs found in the local Universe (Poggianti et al. 2013).
However, our compact galaxies could reflect our initial condition about the sizes of galaxies.
As we describe in \S \ref{gal_model} van der Wel et al. show the variation of galaxy size at high-z.
Thus, we do not conclude that our compact ETGs at z=0 are analogues of those observed 
by Poggianti et al. (2013).

To clarify the differences between the three types of the density profiles, we define the concentrations of insitu and accreted stars by the following equation: $\mathrm{C}=r_{\mathrm{trunc}}/r_{\mathrm{half}}$, where $r_{\mathrm{trunc}} = 60$ kpc given in \S \ref{merger_model} and $r_{\mathrm{half}}$ is the half-mass radius of insitu or accreted stars.
We also define the concentration ratio, $R$, as follows: $R=\mathrm{C_{insitu}} / \mathrm{C_{accreted}}$.
A concentration ratio larger than the unity means that the accreted stars are more extended than the insitu stars.
$\mathrm{C}$ and $R$ of the central galaxies at $z=0$ are summarized in Table \ref{table:concentration_ratio}.
Table \ref{table:concentration_ratio} shows that the ratios $R$ are larger in the first type than in the second type (except H43).
Our results show that the accreted stars in the first type are more extended than those in the second type and contribute to the efficient size growth.
In H43, $R$ is smaller than those in the other first type, and is comparable to those in the second type.
This indicate that H43 has the property intermediate between the first type and the second type.
In the third type, the ratios $R$ are larger than those in other types.
This reflects that the accreted stars in this type contribute only to the outer profile.

\subsection{Surface density profiles}\label{sec:surface_profile}

\begin{figure*}
    \begin{minipage}{180mm}
        \begin{center}
            \begin{tabular}{cc}
                \hspace{-8mm}
                \resizebox{90mm}{!}{\input{./picture/r_log_sigma_log_H07_10.tex}} &
                \hspace{-10mm}
                \resizebox{90mm}{!}{\input{./picture/r_log_sigma_log_H29_08.tex}} \\
            \end{tabular}
            \vspace{-20mm}
            \begin{tabular}{cc}
                \hspace{-8mm}
                \resizebox{90mm}{!}{\input{./picture/r_log_sigma_log_H31_03.tex}} &
                \hspace{-10mm}
                \resizebox{90mm}{!}{\input{./picture/r_log_sigma_log_H33_04.tex}}
            \end{tabular}
            \vspace{-20mm}
            \begin{tabular}{cc}
                \hspace{-8mm}
                \resizebox{90mm}{!}{\input{./picture/r_log_sigma_log_H43_01.tex}} &
                \hspace{-10mm}
                \resizebox{90mm}{!}{\input{./picture/r_log_sigma_log_H44_01.tex}}
            \end{tabular}
            \vspace{-20mm}
            \begin{tabular}{cc}
                \hspace{-8mm}
                \resizebox{90mm}{!}{\input{./picture/r_log_sigma_log_H45_02.tex}} &
                \hspace{-10mm}
                \resizebox{90mm}{!}{\input{./picture/r_log_sigma_log_H48_04.tex}}
            \end{tabular}
            \vspace{-20mm}
            \begin{tabular}{cc}
                \hspace{-93.1mm}
                \resizebox{90mm}{!}{\input{./picture/r_log_sigma_log_H49_05.tex}} \\
            \end{tabular}
            \vspace{-10mm}
            \caption{
            Stellar surface density profiles of the central galaxies in cluster haloes at $z=0$ (thick lines).
            Top left to bottom right: run H07, H29, H31, H33, H43, H44, H45, H48 and H49.
            Thin lines show the initial surface density profiles at $z=2.85$.
            }
            \label{surface_density_profiles}
        \end{center}
    \end{minipage}
\end{figure*}

The stellar surface density profiles of the central galaxies in the nine runs are shown in Fig. \ref{surface_density_profiles}.
Apart from H48 and H49, the growth of the surface densities from their progenitors increases with their radii.
This growth feature is consistent with the `inside-out' growth of ETG surface density profiles reported by \citet{2010ApJ...709.1018V}.
In their observational study, these authors determined the growth of the surface density profile of compact massive ETGs from $z=2$ to $z=0$ using the stacking technique.
The accreted stars increased the surface densities in the outer regions of these galaxies, as shown in \S \ref{sec:density_profile}.
Thus, we find that even major mergers lead to inside-out growth.
In the galaxies of H33 and H45, the surface density increases at $r \la 1$ kpc from their progenitors.
Similar increases in the inner regions have been reported in controlled simulations of dry major mergers (paper I; \citealt{2013MNRAS.429.2924H}).
In H48 and H49, the evolving surface density profiles show no inside-out growth from $z=2.83$, and the surface stellar densities remain within their progenitors, except for small ripple structures in the outer galaxies.

\subsection{Stellar mass--size evolution}\label{sec:mass_size_relation}

\begin{figure*}
    \begin{minipage}{180mm}
        \begin{center}
            \begin{tabular}{cc}
                \hspace{-8mm}
                \resizebox{75mm}{!}{\input{./picture/mass_size_60kpc_z2_201509_1.tex}} &
                \resizebox{75mm}{!}{\input{./picture/mass_size_60kpc_z1_201509_1.tex}} \\
                \hspace{-8mm}
                \resizebox{75mm}{!}{\input{./picture/mass_size_60kpc_z04_201509_1.tex}} &
                \resizebox{75mm}{!}{\input{./picture/mass_size_60kpc_z0_201509_1.tex}}
            \end{tabular}
            \caption{
            Stellar mass-size relations in our simulated cluster galaxies.
            We plot the most massive progenitor galaxies in our simulated cluster haloes.
            The central galaxies are denoted by red empty triangles (H07), red filled circles (H29), green empty triangles (H31), green filled circles (H33), purple filled circles (H43), blue empty triangles (H44), filled brown triangles (H45), olive filled squares (H48) and blue filled squares (H49).
            The solid lines (mean) and brown dotted lines ($\pm 1 \sigma$ of the mean) are the stellar mass--size relations of the local ETGs derived from SDSS (\citealt{2003MNRAS.343..978S}).
            The black dotted lines are the mean stellar mass--size relations of ETGs at $2.0 < z< 2.5$
(\citealt{2012ApJ...746..162N}). 
            Top left to bottom right: $z=2$, $z=1$, $z=0.4$, and $z=0$.        
            (A color version of this figure is available in the online journal.)
            }
            \label{stellar_mass_size_relation}
        \end{center}
    \end{minipage}
\end{figure*}

We now compare the stellar mass--size relations of the central galaxies at various redshifts with those observed in local ETGs.
Figure \ref{stellar_mass_size_relation} plots the size versus mass of the central galaxies at $z=2, 1, 0.4$ and $0$, and the mean local stellar mass--size relations obtained from SDSS (\citealt{2003MNRAS.343..978S}; a similar figure is shown in Figure 8 of \citealt{2013MNRAS.435..901L}).
As an additional comparison, the mean stellar mass--size relations of ETGs in $2.0 < z  < 2.5$, reported by \citealt{2012ApJ...746..162N}, are also plotted.
In our simulations, the central galaxy sizes remain almost constant before $z=2$ because they experience no stellar mergers at these redshifts.
At $z=1$, all central galaxies remain below the locally obtained plots.
At $z=0.4$, four galaxies are within $\pm 1\sigma$ of the mean local plot.
We have confirmed that for most of the central galaxies the masses and sizes increase by dry major and minor mergers at $z \le 1$.
The mass growth is consistent with the numerical results from \citet{2013MNRAS.435..901L} as well as the observational results from \citet{2008MNRAS.385...23L, 2009MNRAS.396.2003L}, \citet{2013MNRAS.433..825L}, and \citet{2013MNRAS.434.2856B}.
At $z=0$, roughly 60\% of the central galaxies lie within $\pm 1\sigma$ of the mean local plot.
Therefore, dry major and minor mergers are sufficient to increase the masses and sizes of the simulated compact massive ETGs to those of local ETGs.
Our results agree with other numerical studies which focused the size growth of compact massive ETGs from $z\sim2$ to $z=0$ at different mass ranges and using different 
methods (\citealt{2009ApJ...699L.178N}; \citealt{2012ApJ...744...63O}; \citealt{2013MNRAS.435..901L}).
The central galaxies in H48 and H49 remain far below the local plot because their merger events are too small to alter the central galaxies (see \S \ref{sec:merger_histories}).

\begin{figure}
                \resizebox{80mm}{!}{\input{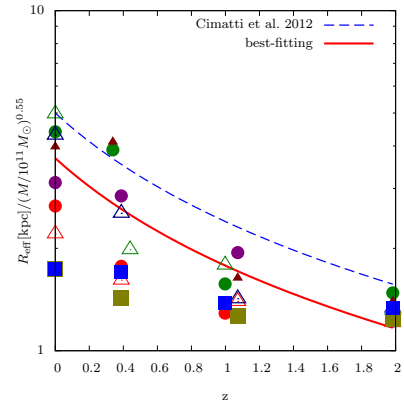}} \\
  \caption{Size evolution of our simulated galaxies as a function of redshift.
  The vertical axis shows the mass-normalized effective radius, $R_{eff}/M^{0.55}$.
  The blue dashed line is the observational result of \citet{2012MNRAS.422L..62C}, approximated as $ R_{eff}/M^{0.55} \propto (1+z)^{-1.06}$.
  The red solid line is the average size evolution of the central galaxies in our numerical results, approximated as $R_{eff}/M^{0.55} \propto (1+z)^{-1.09}$.
  In this analysis, H48 and H49 are excluded because their galaxies remain roughly compact from $z=2.85$.
  The colored symbols are explained in the caption to Fig. \ref{stellar_mass_size_relation}.
  (A color version of this figure is available in the online journal.)
  }
  \label{z_size}
\end{figure}

Figure \ref{z_size} plots the redshift evolution of the central galaxy sizes.
For comparison with the observations of \citet{2012MNRAS.422L..62C}, we plot the mass-normalized effective radius $R_{\mathrm{eff}}/M^{0.55}$ versus the redshift.
Excluding H48 and H49, the average size of the central galaxies evolves as $R_{\mathrm{eff}}/M^{0.55} \propto (1+z)^{-1.09}$.
The trend of the average size evolution is similar to that of \citet{2012MNRAS.422L..62C}.
Our sample is, however, too scarce to compare our result with the \citet{2012MNRAS.422L..62C} quantitatively, which exploited  a large sample of 1975 ETGs at $0.2<z<3$ for their analysis.

\subsection{Stellar-to-halo mass relation}

\begin{figure}
    \resizebox{88mm}{!}{\input{./picture/m_dm_m_star_z0_SHM_201509_6.tex}}    
  \caption{Stellar-to-halo mass relations of the central galaxies at $z=0$.
  The dashed-dotted line was obtained by the abundance matching method, performed at $z=0$ by \citet{2010ApJ...710..903M}.
  The thick solid, dashed, and dotted lines are the results of \citet{2013MNRAS.428.3121M}, \citet{2013ApJ...770...57B}, and \citet{2013ApJ...771...30R}, respectively.
  The dark gray area and the error bars represent the plausibility range and the 1$\sigma$ scatter bars around the relation of \citet{2013MNRAS.428.3121M} and \citet{2013ApJ...770...57B}.
  The colored symbols are explained in the caption to Fig. \ref{stellar_mass_size_relation}.
  (A color version of this figure is available in the online journal.)
  }
  \label{m_dm_m_star_z0}
\end{figure}

Finally, we compare the simulated and observationally inferred stellar-to-halo mass relations in galaxies at $z=0$.
Figure \ref{m_dm_m_star_z0} shows our simulated results and the stellar-to-halo mass relations obtained by abundance matching at $z=0$ by \citet{2010ApJ...710..903M}, \citet{2013MNRAS.428.3121M}, \citet{2013ApJ...770...57B}, and \citet{2013ApJ...771...30R}.
The inferred mass ratios of stellar to dark matter components are lower in massive dark matter haloes than in less massive ones (\citealt{2010ApJ...710..903M}; \citealt{2010MNRAS.404.1111G}; \citealt{2010ApJ...717..379B}).
Our simulations qualitatively replicate these results.
In contrast, cosmological hydrodynamic simulations overproduce stars in cluster-sized haloes by over-cooling their gas content (\citealt{2012MNRAS.425..641L}; \citealt{2013MNRAS.433.3297D}; \citealt{2013MNRAS.436.1750R}).
Our present model simulates stellar and dark matter components but excludes star formation.
Essentially, we simulate star formation quenching in all dark matter haloes at $z=2.85$.
All our simulated galaxies are roughly consistent with  \citet{2013ApJ...770...57B} and \citet{2013ApJ...771...30R}, although they lie at $\sim$0.2--0.4 dex below the average stellar-to-halo mass relation at $z=0$ reported by \citet{2010ApJ...710..903M, 2013MNRAS.428.3121M}.

\section{Summary and Discussion}
Using a re-simulation technique, we demonstrated that central galaxies in cluster haloes of masses 1--5 $\times 10^{14}\mathrm{M}_{\odot}$ double their average mass and quadruple their size from $z=2$ to $z=0$ by dry mergers.
The simulated growths favorably agree with \citet{2010ApJ...709.1018V}, who conducted a stacking analysis of the observational data, and \citet{2015ApJ...802...73S}, who performed a detailed comparison between their semi-empirical model and observational data.
By simulating galaxy merger processes in high density environments, we showed that major mergers are important contributors to the mass growth of the central galaxies of 1--5 $\times 10^{14}\mathrm{M}_{\odot}$ cluster haloes (see Table \ref{table:merger_mass_ratio}).
The central galaxies (except H48 and H49) experienced one or two dry major mergers because the number density of the massive galaxies exceed the Universe mean under such environments.
Our result also agrees with the previous studies which investigated the size growth of compact massive ETGs at the different halo mass ranges (\citealt{2009ApJ...699L.178N}; \citealt{2012ApJ...744...63O}; \citealt{2013MNRAS.435..901L}), and confirms the evolution of ETGs driven by dry major and minor mergers.

The stellar density and surface density profiles of our simulated central galaxies exhibited inside-out growth from $z=2.85$ to $z=0$.
This inside-out growth is consistent with the observational analysis of \citet{2010ApJ...709.1018V}.
The exceptions were runs H48 and H49, in which the central galaxies were as compact and massive as their progenitors at $z=2.85$ and experienced only a few minor merger events.

According to our numerical results, the central galaxies in dark matter haloes satisfying $5\times 10^{14} \mathrm{M}_{\odot}\ga M_{\mathrm{halo}} \ga 10^{14} \mathrm{M}_{\odot}$ are potential descendants of high-z ($z\sim 2 - 3$) compact massive ETGs, as argued by the other studies at the different mass ranges  (\citealt{2009ApJ...699L.178N}; \citealt{2012ApJ...744...63O}; \citealt{2013MNRAS.435..901L}).
The relationship between the stellar mass of the central galaxies and the mass of the cluster haloes at $z=0$ roughly agrees with the observational results of \citet{2013ApJ...770...57B} and \citet{2013ApJ...771...30R}.

In most of our simulated central galaxies at $z=0$, the mass--size relation is similar to that of local ETGs reported by \citet{2003MNRAS.343..978S}.
Moreover, the redshift evolution of the average mass-normalized size of the central galaxies approximates the observations of \citet{2012MNRAS.422L..62C}, although two of the central galaxies remain compact in our simulations.

The central galaxies of cluster haloes appear to enlarge and accrete mass by dry major mergers (see Table \ref{table:size_growth_efficiency} and \ref{table:merger_mass_ratio}).
However, major mergers have been considered to be ineffectual for mass--size growth by the following simple argument (see \citealt{2009ApJ...697.1290B}).
In major mergers of two masses $M_1$ and $M_2$ in parabolic orbits, energy conservation gives $R\propto M$ as follows:
\begin{equation}
\frac {GM^2}{R}=\frac{GM_1^2}{R_1}+\frac{GM_2^2}{R_2};
\end{equation}
and provided that the two progenitor galaxies and the merger remnant are structurally homologous, and provided that $M=M_1+M_2$ by mass conservation.
Then, equal mass mergers ($M_1=M_2$) result in $R=2R_1$.
By this mechanism, major mergers cannot explain the observed mass--size evolution of ETGs, which satisfies $R_e\propto M_* ^{\alpha}$ with $\alpha \sim 2$ (\citealt{2009ApJ...697.1290B}; \citealt{2010ApJ...709.1018V}).
Rather, our numerical results show that major mergers can lead to inside-out growth of ETGs in cluster haloes, which violates the structural homology assumption.
Such inside-out growth can be explained by the small orbital angular momenta of the merging galaxies in our cosmological simulation.
As shown by \citet*{2006MNRAS.369.1081B}, if major mergers possess low angular momentum, they lose less energy through dynamical friction with the dark matter of central galaxies than mergers with substantial angular momentum; consequently, they can leave extended remnants.
Therefore, we argue that major mergers can effectively contribute to galactic mass--size growth.

Our result agrees with the other observational studies.
\citet{2012A&A...548A...7L} analysed massive ETGs in the COSMOS field, and showed that nearly half of the galaxies with stellar masses exceeding $10^{11} \mathrm{M}_{\odot}$ acquired their extra mass through major mergers ($M_2 / M_1 \geq 1/4$ at $z \la 1$).
\citet{2014MNRAS.444..906F} formed a similar conclusion from the SHARDS survey.
The importance of major mergers has also been highlighted in the brightest cluster galaxies (\citealt{2008MNRAS.385...23L, 2009MNRAS.396.2003L}; \citealt{2013MNRAS.433..825L}; \citealt{2013MNRAS.434.2856B}).
Together with our results, these observations show that major mergers are important contributors to the mass growth of massive galaxies.

Recently, \citet{2014arXiv1409.1924L} performed $N$-body simulations which explicitly followed the evolution of both stars and dark matter in clusters.
They showed that the evolution of their brightest cluster galaxies did not differ appreciably from those found in the simulations which are used the weighting scheme.
This last study notwithstanding, it is worth investigating the mass and size growth using the re-simulation technique which we have performed.

We note that our simulations have a number of limitations.
For initial structures of the galaxies, we assume that all galaxies are compact and follow 
the stellar mass-size relation of ETGs at $2.0<z<2.5$ obtained by \citet{2012ApJ...746..162N}.
However, \citet{2014ApJ...788...28V} show the intrinsic scatter in size in a given mass range
and not all $z\sim2$ galaxies are compact.
In addition, \citet{2013ApJ...777...18M} showed that the passive fraction is about 30 per cent for massive galaxies.
Thus, our initial condition could be too simplistic to represent high-z galaxies.
We should consider the structural variation of progenitor galaxies in future studies.
Furthermore, in our simulations, we ignored additional star formation.
However, this could contribute to mass growth from z=3 to z=0 (e.g. \citealt{2014MNRAS.445.2198O}).
Therefore, we should incorporate star formation as an extension of our present galaxy formation approach.

\section*{Acknowledgments}

We are grateful to the referee for providing for constructive comments that improved the paper.
We thank Volker Springel for making GADGET-2 publicly available.
We thank Takashi Okamoto and Masahiro Nagashima for their fruitful discussions and comments.
Numerical computations were partially carried out on Aterui and Cray XT4 supercomputers at Center for Computational Astrophysics, CfCA, of National Astronomical Observatory of Japan, and the K computer at the RIKEN Advanced Institute for Computational Science (Proposal numbers hp120286).
This work is partly supported by Grant-in-Aid for COE Research (20001003) of  Japan Society for the Promotion of Science (to A.H.).
This study has been funded by MEXT/JSPS KAKENHI Grant Number 25287041.
TI has been funded by MEXT HPCI STRATEGIC PROGRAM, MEXT/JSPS KAKENHI Grant Numbers 24740115 and 15K12031.

\bibliographystyle{mn}
\bibliography{mn-jour,ref}


\bsp

\appendix

\section{Density profiles of the merger remnant of an  equal mass dry major merger}\label{sec:appendix}

\begin{figure}
                \resizebox{75mm}{!}{\input{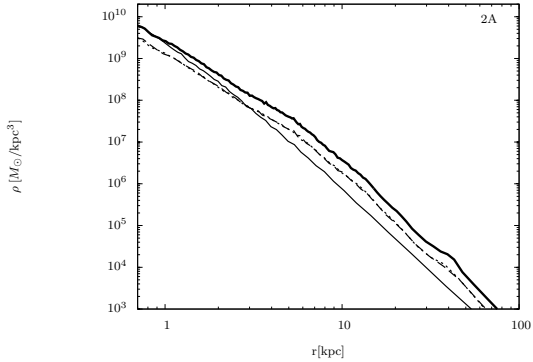}}
            \caption{
            Stellar density profiles of the remnant stellar system of a dry major merger (run 2A in paper I).
            As in Fig. \ref{density_profiles}, the thick (thin) line is the density profile of all stars in the merger remnant (the initial galaxy model).
            The dotted and dashed lines denote the stars belonging to the two progenitor galaxies involved in the merger.
            }
            \label{density_profiles_2A}
\end{figure}

Figure \ref{density_profiles_2A} shows the stellar density profiles of the merger remnant of an equal mass dry major merger (run 2A in paper I).
In paper I, the major merger was assumed as a parabolic, head-on encounter of two equal-mass galaxies (i.e. a mass ratio of 1:1).
Moreover, the galaxies were assumed to be stellar bulges with dark matter haloes.
The initial model and the simulation parameters are detailed in paper I.
From Fig. \ref{density_profiles_2A}, we observe that the stars in both progenitor galaxies (approximately) equally contribute to the stellar density of the remnant at all radii.
In other words, both progenitors make almost equal contributions to the remnant profile.
This behavior significantly differs from that of dry minor mergers (see paper I), especially in the inner region ($r < 1$ kpc).
The mass--size growth in Fig. \ref{density_profiles_2A} is given by $R_e\propto M_* ^{\alpha}$, where $\alpha \sim 1.6$ (paper I). 

\label{lastpage}
\end{document}